*Research Article*

# Cross Section Prediction for Inclusive Production of $Z$ Boson in pp Collisions at $\sqrt{s} = 14$ TeV: A Study of Systematic Uncertainty due to Scale Dependence

**Hasan Ogul[1,2] and Kamuran Dilsiz[2,3]**

[1]*Department of Nuclear Engineering, Sinop University, 57000 Sinop, Turkey*
[2]*Department of Physics and Astronomy, University of Iowa, Iowa City, IA 52242, USA*
[3]*Department of Physics, Bingöl University, 12000 Bingöl, Turkey*

Correspondence should be addressed to Hasan Ogul; hsnogul@gmail.com

Prediction of $Z \to l^+l^-$ production cross section (where $l^{\pm} = e^{\pm}, \mu^{\pm}$) in proton-proton collisions at $\sqrt{s} = 14$ TeV is estimated up to next-to-next-to-leading order (NNLO) in perturbative QCD including next-to-leading order (NLO) electroweak (EW) corrections. The total inclusive $Z$ boson production cross section times leptonic branching ratio, within the invariant mass window $66 < m_{ll} < 116$ GeV, is predicted using NNLO HERAPDF2.0 at NNLO QCD and NLO EW as $\sigma_Z^{\text{Tot}} = 2111.69^{+26.31}_{-26.92}$ (PDF) $\pm 11$ ($\alpha_s$) $\pm 17$ (scale) $^{+57.41}_{-30.98}$ (parameterization and model). Theoretical prediction of the fiducial cross section is further computed with the latest modern PDF models (CT14, MMHT2014, NNPDF3.0, HERAPDF2.0, and ABM12) at NNLO for QCD and NLO for EW. The central values of the predictions are based on DYNNLO 1.5 program and the uncertainties are extracted using FEWZ 3.1 program. In addition, the cross section is also calculated as functions of $\mu_R$ and $\mu_F$ scales. The choice of $\mu_R$ and $\mu_F$ for scale variation uncertainty is further discussed in detail.

## 1. Introduction

The Large Hadron Collider (LHC) has played a crucial role in recent particle physics discoveries and is currently the main tool for exploring TeV scale physics. The LHC is presently the biggest and most powerful proton-proton (pp) collider and allows to test Standard Model (SM) predictions in different center of mass ($\sqrt{s}$) energies. The LHC started to run with a collision energy of 7 TeV in 2010 and the energy was raised to 8 TeV in 2012. The time period between 2010 and 2012 is called Run I. Then, there was a break between 2012 and 2015 at the LHC for the maintenance and upgrade of the collider. The LHC restarted to run at $\sqrt{s} = 13$ TeV (Run II) early in 2015 in order to optimize the delivery of particle collisions for physics research. The second run of the LHC saw an unprecedented data set collected in 2015 and 2016 with more expected in 2017 which opens new possibilities for searches for new physics and precision measurements. The collider was designed to run at a maximum collision energy of 14 TeV; therefore, 14 TeV predictions are considered in this paper. More details about the LHC and its schedule could be found in [1].

Electroweak boson production at LHC provides a major testing ground for quantum chromodynamic (QCD) and electroweak (EW) processes. $Z$ boson production is one of the most prominent examples of electroweak events. It has large cross section and clean experimental signature, which allow for important precision tests of SM. The process is also very valuable for the calibration of the detector [2, 3] and provides substantial input for the parton distribution function (PDF) determination [4–6].

Inclusive $Z$ boson production cross section was previously measured by the ATLAS and CMS Collaborations at the LHC at $\sqrt{s} = 7$ TeV [7, 8], $\sqrt{s} = 8$ TeV [9], and $\sqrt{s} = 13$ TeV [10–12]. ATLAS and CMS publications showed that there is a good agreement between the observed and NNLO predicted results. Therefore, the theoretical predictions of higher-order corrections are essential work to validate the experimental results.

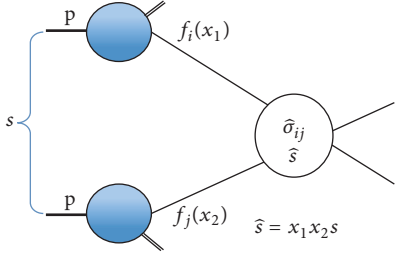

Figure 1: Hard parton scattering producing weak bosons.

Theoretical prediction of $Z$ boson is available up to next-to-next-to-leading order (NNLO) in perturbative QCD and to next-to-leading order (NLO) for EW processes. The cross section definition of a weak boson in pp collisions is given by (1), and the process is illustrated in Figure 1. Here $i$, $j$ are the incoming partons with the momentum fraction $x_{i,j}$, $f_{i,j}(x_{1,2})$ represent PDFs, and $\hat{\sigma}_{i,j}$ stands for partonic cross section. PDFs are known with limited precision, and partonic cross section is calculable to high orders (QCD, EW). Total and fiducial cross section predictions of $Z \rightarrow l^+l^-$ events in pp collisions are presented providing PDF, strong coupling ($\alpha_s$), scale, model, and parameterization uncertainties based on DYNNLO 1.5 [13] and FEWZ 3.1 [14]. The central values of the predictions are calculated using DYNNLO 1.5, and the theoretical uncertainties and the impact of NLO EW correction on the cross section are extracted using FEWZ 3.1 MC generator program. The disagreement between FEWZ 3.1 and DYNNLO 1.5 is previously reported as smaller than 1% [15]. The selections used in this analysis are the same as those used in the latest ATLAS study at 13 TeV [10]. The total and fiducial cross section are estimated up to NNLO QCD including NLO EW corrections with the most recent PDF sets (HERAPDF2.0 [16], CT14 [17], NNPDF 3.0 [18], MMHT 2014 [19], and ABM12 [20]).

$$\sigma(pp \longrightarrow X) = \sum_{i,j} \iint dx_1 dx_2 f_i(x_1, \mu_F) f_j(x_2, \mu_F) \hat{\sigma}_{i,j}(x_1 x_2 s, \mu_R). \quad (1)$$

Another topic investigated in this paper is the parameter choices for scale uncertainty. The scale uncertainty depends on the choice of renormalization ($\mu_R$) and factorization ($\mu_F$) scales. Therefore, two different approaches for scale uncertainty calculation are also studied using HERAPDF2.0, as well as the cross section dependence of $\mu_R$ and $\mu_F$ scales used in the perturbative calculation. The standard convention of considering the range $M_Z/2 < \mu < 2M_Z$ is followed, setting $\mu_R = \mu_F = \mu$. Besides the standard convention, the possibility of varying independently the value of renormalization and factorization scale is investigated, and the main findings are discussed.

## 2. Total and Fiducial Cross Section

The total inclusive $Z$ boson production cross section times leptonic branching ratio, within the invariant mass window $66 < m_{ll} < 116$ GeV, is estimated at $\sqrt{s} = 14$ TeV in pp collisions. A fiducial region is further defined and $Z$ boson production cross section is computed. For the fiducial cross section, the leptons are required to have $|\eta| < 2.5$ and $p_T > 25$ GeV. The invariant mass window is kept as $66 < m_{ll} < 116$ GeV. All presented results are for born level (pre-QED FSR) leptons. HERAPDF2.0, CT14, NNPDF3.0, MMHT2014, and ABM12 PDF models are used with the corresponding value of $\alpha_s(M_Z) = 0.118$ and $M_Z = 91.1876$ GeV. For completeness, we first compute the predictions for 13 TeV and the predicted values are compared with ATLAS collaboration result [10]. Then, we rerun the code for 14 TeV center of mass energy.

Figure 2 presents the 13 TeV and 14 TeV predictions we performed. The upper two figures prove that there is a good agreement between experimental measurements and theoretical predictions. Particularly, HERAPDF2.0 prediction at NNLO best describes the measurement. The solid thick lines on upper figures represent the central value of ATLAS measurement. The dashed and dotted lines show the quadrature sum of the luminosity and systematic uncertainties and only systematic uncertainty, respectively. Black points stand for the central values of the predictions. The green bands present only scale uncertainty and the yellow bands illustrate the quadrature sum of PDF, scale, and $\alpha_s$ for MMHT2014, NNPDF3.0, CT14, and ABM12 PDF predictions if they are available. The green band for HERAPDF2.0 is the quadrature sum of PDF, scale, $\alpha_s$, parameterization, and model uncertainties.

The PDF uncertainty is estimated following closely the prescription of PDF4LHC working group [21, 22]. These uncertainties are evaluated with five different NNLO PDFs for the fiducial cross section: HERAPDF2.0, CT14, NNPDF3.0, MMHT2014, and ABM12. The PDF uncertainty of the CT14 PDF set is rescaled from 90% CL to 68% CL to match the other sets. The PDF uncertainty of the total cross section is estimated using LO, NLO, and NNLO HERAPDF2.0 PDF sets at LO, NLO, and NNLO QCD, respectively. The $\alpha_s$ uncertainty is estimated varying $\alpha_s$ by ±0.001. Then, the uncertainties are symmetrized by taking the bigger value from estimated up and down errors. The scale uncertainties are calculated varying the value of $\mu_R$ and $\mu_F$ scales. The default value of $\mu_{R,F}$ is set to $M_Z$, which is taken as 91.1876 GeV. The scales are changed by factors of two, $M_Z/2 \leq \mu \leq 2M_Z$, where $\mu_R = \mu_F = \mu$. The maximum value of the variation is taken as the scale uncertainty, which makes the scale uncertainty symmetric. To calculate the parameterization and model uncertainties, HERAPDF20_NLO_VAR and HERAPDF20_NNLO_VAR PDF sets are used. Each PDF set has 13 eigenvectors and each variation is treated one by one to estimate the model and parameterization errors. First, model and parameterization errors are calculated separately, and, then, they are summed in quadrature. More details about the calculation of model and parameterization errors can be found in [23]. After the calculation of all considered uncertainties, the total uncertainty is determined by adding the individual uncertainties in quadrature. The numerical values are provided in Results.

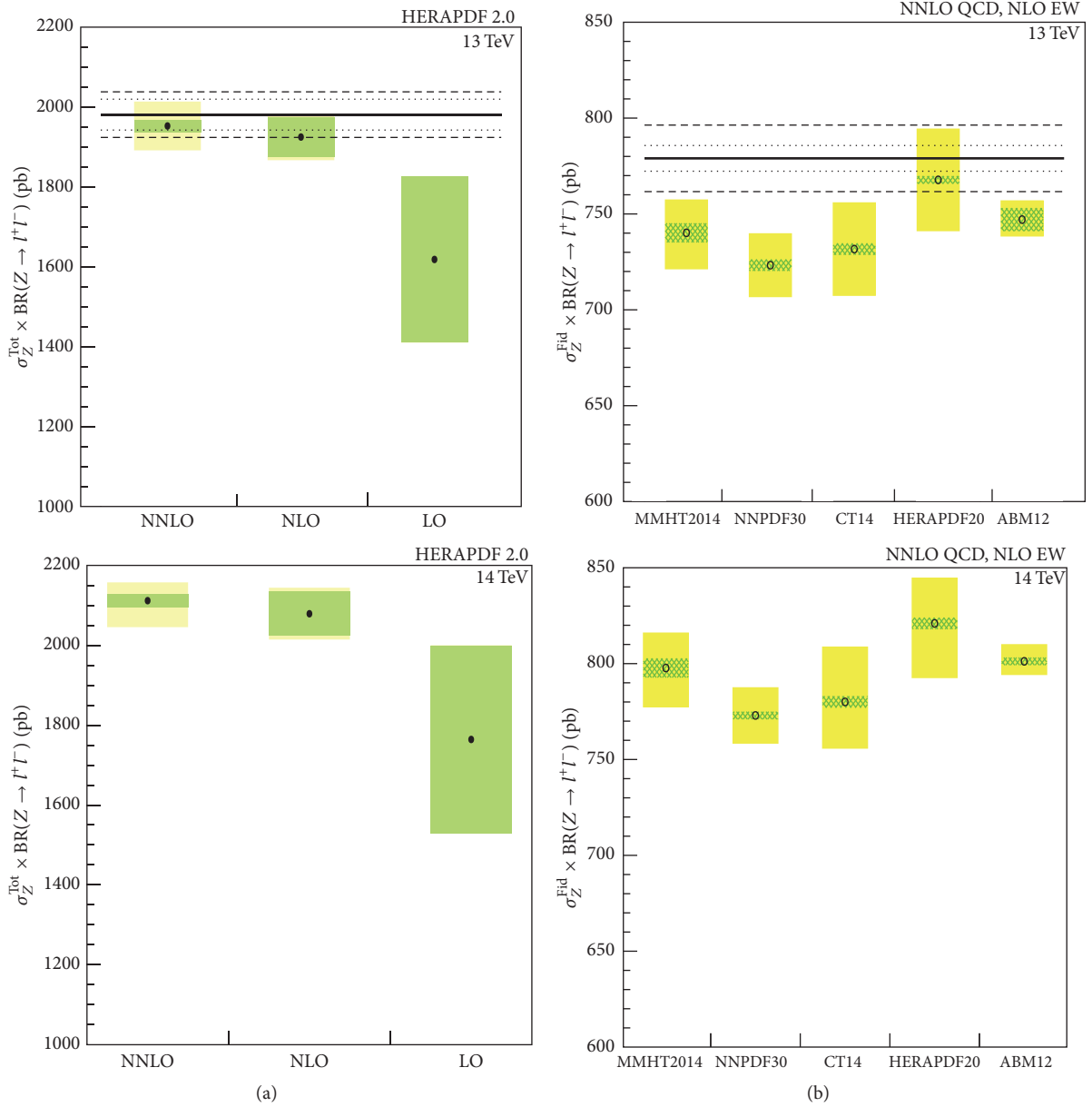

Figure 2: The total (a) and fiducial (b) inclusive $Z$ boson production cross section times leptonic branching ratio. The upper plots show 13 TeV results and the lower ones present 14 TeV QCD predictions.

## 3. Renormalization and Factorization Scales

The source of the renormalization scale is ultraviolet divergence, which is one of the most important aspects of field theories, and the factorization scale arises from collinear (infrared) divergence. Proper treatment of these divergences is crucial to understand the physics processes and perturbation theory. Any wrong choice of $\mu_R$ and $\mu_F$ scales may lead to an unstable perturbative behavior. The details and definition of the divergences and the scales can be found in a well written paper [24].

The hadronic cross section production for Drell-Yan production or other LHC processes is already defined in (1). The partonic cross section ($\hat{\sigma}_{i,j}(x_1 x_2 s, \mu_R)$) can be written in terms of LO, NLO, and NNLO as follows:

$$\hat{\sigma}_{pp \to Z} = \left[ \hat{\sigma}_{LO} + \alpha_s(\mu_R) \hat{\sigma}_{NLO} + \alpha_s^2(\mu_R) \hat{\sigma}_{NNLO} + \cdots \right]_{pp \to Z}. \quad (2)$$

Table 1: Cross section variations with regard to the central value of the prediction based on $\mu_R$ and $\mu_F$ scales at NLO and NNLO QCD. All numbers are in units of pb.

| $(\mu_R, \mu_F)$ | (1, 1) | (1, 2) | (1, 1/2) | (2, 1) | (2, 2) | (2, 1/2) | (1/2, 1) | (1/2, 2) | (1/2, 1/2) |
|---|---|---|---|---|---|---|---|---|---|
| NNLO | 0 | 7 | 20 | 1 | 17 | 35 | 6 | 10 | 14 |
| NLO | 0 | 64 | 101 | 24 | 54 | 138 | 29 | 18 | 55 |

The cross section statement turns into (3), which is the cross section definition in terms of LO, NLO, NNLO, and so on.

$$\sigma(\text{pp} \longrightarrow Z) = \iint dx_1 dx_2 f_1(x_1, \mu_F) f_2(x_2, \mu_F) \cdot \left[\hat{\sigma}_{LO} + \alpha_s(\mu_R) \hat{\sigma}_{NLO} + \alpha_s^2(\mu_R) \hat{\sigma}_{NNLO} + \cdots\right]. \quad (3)$$

We may also express the cross section in terms of analytical expression of these scales.

$$\sigma(x_1 x_2, Q^2) = \sum_{i,j} \int_0^1 \frac{dx_1}{x_1} \frac{dx_2}{x_2} f_i(x_1, \mu_F^2) f_j(x_2, \mu_F^2) \cdot G_{ij}\left(x_1 x_2; \alpha_s(\mu_R^2), \frac{Q^2}{\mu_R^2}; \frac{Q^2}{\mu_F^2}\right), \quad (4)$$

where $Q^2$ is the experimental energy scale and it is set to $M_Z^2$ in our case. The hard scattering function $G_{ij}$ is presented as an analytical expression of $\mu_R$ and $\mu_F$ and it can be further expanded as follows:

$$G_{ij}\left(x_1 x_2; \alpha_s(\mu_R^2), \frac{M_Z^2}{\mu_R^2}; \frac{M_Z^2}{\mu_F^2}\right)$$

$$= \sum_{n=0}^{+\infty} \left(\frac{\alpha_s^2(\mu_R^2)}{\pi}\right)^n G_{ij}\left(x_1 x_2; \frac{M_Z^2}{\mu_R^2}; \frac{M_Z^2}{\mu_F^2}\right)$$

$$= G_{ij}^0(x_1 x_2) + \frac{\alpha_s(\mu_R^2)}{\pi} G_{ij}^1\left(x_1 x_2; \frac{M_Z^2}{\mu_R^2}; \frac{M_Z^2}{\mu_F^2}\right) \quad (5)$$

$$+ \frac{\alpha_s^2(\mu_R^2)}{\pi^2} G_{ij}^2\left(x_1 x_2; \frac{M_Z^2}{\mu_R^2}; \frac{M_Z^2}{\mu_F^2}\right) + \mathcal{O}(\alpha_s^3),$$

where $G_{ij}^0(x_1 x_2)$, $G_{ij}^1(x_1 x_2)$, and $G_{ij}^2(x_1 x_2)$ terms give the LO, NLO, and NNLO contributions, respectively. $\mathcal{O}(\alpha_s^3)$ term represents the contributions beyond NNLO calculations. FEWZ and DYNNLO programs let us set the factorization and renormalization scales different to calculate the cross sections, so they successfully serve the purpose of our paper.

Figure 3 shows the total cross section dependence on $\mu_R$ and $\mu_F$ scales. The values are estimated at LO, NLO, and NNLO QCD using DYNNLO 1.5 MC generator program interfaced with LO, NLO, and NNLO HERAPDF2.0 PDF sets. $\alpha_s$ is a renormalization scale dependent parameter and does not appear at leading order since LO $Z$ boson diagram is purely electroweak. Based on (3), it starts to take a role at NLO. This feature can be proven by plotting total cross section as a function of $\mu_R$. Figure 3(b) presents LO, NLO, and NNLO total cross section predictions as a function of $\mu_R$.

Here $\mu_F$ is fixed to $M_Z$ while $\mu_R$ is varied. It is clearly seen that LO QCD prediction is constant for different $\mu_R$ values. NLO and NNLO QCD predictions decrease by increase of $\mu_R$. NNLO prediction is always higher than NLO prediction. In Figure 3(a), $\mu_R$ and $\mu_F$ scales are varied by setting the condition: $\mu_R = \mu_F$. The cross section increases by increase of $\mu_R$ and $\mu_F$. The increase is the least for NNLO QCD prediction and the most for LO. After some points, the LO and NLO QCD predictions become higher than NNLO QCD prediction. As can be seen from (3), PDFs are dependent on $\mu_F$ parameter, and $\mu_F$ scale starts to contribute the calculations at LO. Figure 3(c) illustrates the cross section variation by $\mu_F$. Here $\mu_R$ is set to $M_Z$ while $\mu_F$ varies. The findings are similar to the one observed for the variation of scales with condition of $\mu_R = \mu_F$.

There are two approaches to calculate the scale uncertainty. The first one is the standard convention researchers often used. It is considered that $\mu$ is multiplied by half or 2, where $\mu_R = \mu_F = \mu$. Then, the difference with the default value is taken as the scale uncertainty (see (6)). The second way is varying independently the value of renormalization and factorization scales. The maximum variation is assigned as scale uncertainty (see (7)).

$$\Delta\text{Scale}_1 = \max\left(\sigma_{\mu_{R,F}=M_Z} - \sigma_{\mu_{R,F}=M_Z/2}, \sigma_{\mu_{R,F}=2M_Z} - \sigma_{\mu_{R,F}=M_Z}\right), \quad (6)$$

$$\Delta\text{Scale}_2 = \max\left(\sigma_{\mu_{R,F}=M_Z} - \sigma_{\mu_R=mM_Z, \mu_F=nM_Z}\right), \quad (7)$$

where $m$ and $n$ could be 1/2, 1, or 2. We have total 9 variations including the default one ($m = 1, n = 1$). Those are listed as (1, 1), (1, 2), (1, 1/2), (2, 1), (2, 2), (2, 1/2), (1/2, 1), (1/2, 2), and (1/2, 1/2). The scale uncertainty calculated at NLO and NNLO QCD is illustrated by Figure 4. The absolute value of each variation is also given in Table 1. The renormalization and factorization scales have different origins as stated above. Therefore, they should be varied independently by a factor of two. It is however found that large logarithms arise when the two scales differ by a factor of 4, and this gives spuriously large uncertainties, so these special configurations, that is, $\mu_R/\mu_F = 4$, should be removed from the calculation of the scale uncertainties. For example, at NNLO QCD, the scale uncertainty for two approaches is as follows: $\Delta\text{Scale}_1 = 17$ and $\Delta\text{Scale}_2 = 35$. The uncertainty with $\Delta\text{Scale}_2$ is about two times bigger than the standard convention method. At NLO QCD, the scale uncertainty for two approaches is as follows: $\Delta\text{Scale}_1 = 55$ and $\Delta\text{Scale}_2 = 138$. This time, $\Delta\text{Scale}_2$ is almost three times bigger than $\Delta\text{Scale}_1$.

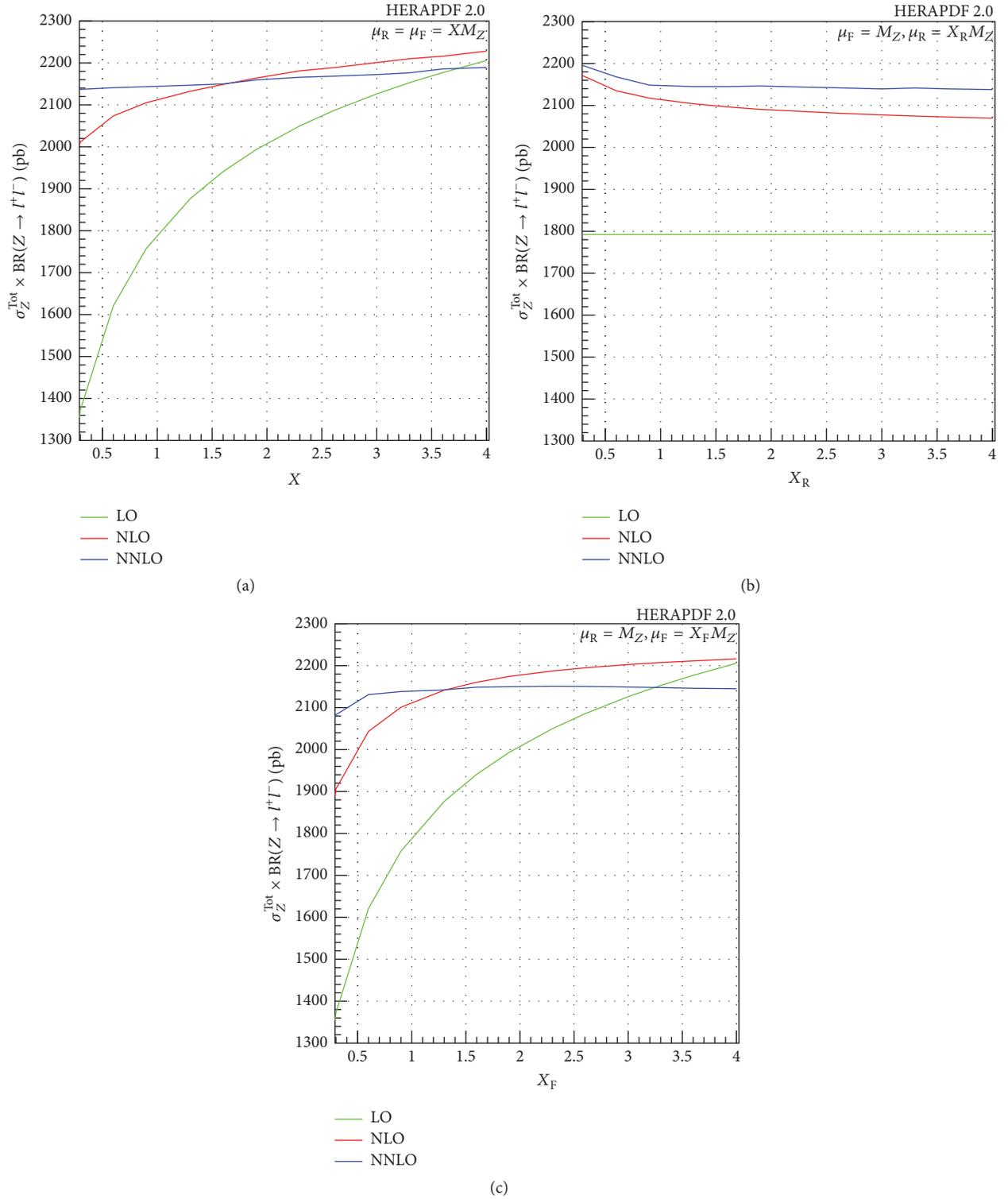

Figure 3: $\mu_R$ and $\mu_F$ dependence of total LHC cross section at $\sqrt{s} = 14$ TeV for $Z \to l^+l^-$ production at LO, NLO, and NNLO QCD.

## 4. Results

The total and fiducial cross sections of inclusive $Z$ boson production are given in Tables 2 and 3. The numbers are obtained with LO, NLO, and NNLO QCD and NLO EW corrections as stated before. Here the total uncertainty is the quadrature sum of the PDF, $\alpha_s$, scale, model, and parameterization errors. All are calculable for NNLO and NLO HERAPDF2.0 PDF sets but model and parameterization errors are not available for LO HERAPDF2.0 set. PDF, $\alpha_s$ and scale uncertainties are calculable for NNLO MMHT2014, NNPDF3.0, and CT14 PDF sets. Only PDF and scale uncertainties are available for NNLO ABM12 PDF set.

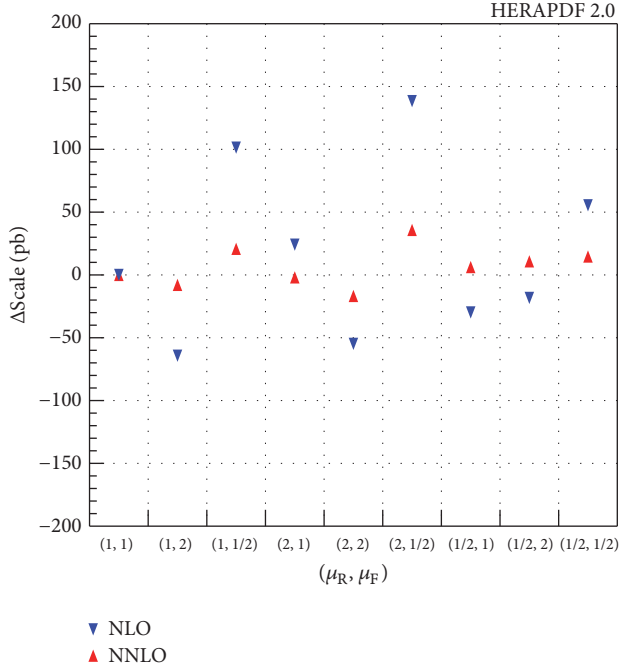

Figure 4: Scale uncertainty with different factorization and renormalization choices at NLO and NNLO QCD.

Table 2: Predicted 14 TeV total cross section values at LO, NLO, and NNLO QCD and NLO EW. All numbers are in units of pb.

| | HERAPDF2.0 | | |
|---|---|---|---|
| | Cross section | Total unc. | Scale unc. |
| NNLO QCD + NLO EW | 2111.69 | +66.56 −45.41 | ±17 |
| NLO QCD + NLO EW | 2078.94 | +64.60 −65.13 | ±55 |
| LO QCD + NLO EW | 1764.20 | +235.80 −235.98 | ±235 |

Table 3: Predicted 14 TeV fiducial cross section values at NNLO. All numbers are in units of pb.

| | NNLO QCD + NLO EW | | |
|---|---|---|---|
| | Cross section | Total unc. | Scale unc. |
| HERAPDF2.0 | 820.96 | +28.22 −23.59 | ±3 |
| MMHT2014 | 797.71 | +20.21 −18.20 | ±5 |
| NNPDF3.0 | 772.96 | +14.41 −14.41 | ±2 |
| CT14 | 780.03 | +24.09 −28.57 | ±3 |
| ABM12 | 801.19 | +6.82 −8.67 | ±2 |

The cross section values of full phase-space are presented in Table 2. Based on these numerical values, it can be safely stated that the scale uncertainty is getting smaller from LO to NNLO QCD as expected. The scale uncertainty at NNLO is 25.75% of the calculated total uncertainty and the scale uncertainty contributions to the total uncertainty at NLO and LO are 84.61% and 99.58%, respectively. Table 3 provides the fiducial cross section numbers of inclusive production of $Z$ boson in pp collisions at NNLO QCD and NLO EW. These numbers are produced using five different PDF models: HERAPDF2.0, MMHT2014, NNPDF3.0, CT14, and ABM12. The table proves that the predictions based on different PDF models are consistent with each other. The CT14 prediction has the biggest total uncertainty. The MMHT2014 prediction has the biggest scale uncertainty but overall all PDF models have the same level of scale uncertainties. ABM12 has the smallest total uncertainty since only PDF and scale uncertainties are calculable for ABM12 predictions. As can be seen from these numbers, the scale parameters are crucial for the QCD calculations.

The numerical values of cross sections based on variation of $\mu_R$ and $\mu_F$ are given in Table 4. The table also provides the cross section ratios at different orders, which is so-called $k$-Factor. Table 4 is already illustrated in Figure 3. In Figure 3(a), LO prediction has the sharpest slope while NNLO prediction has the smallest slope. This illustrates why the scale uncertainty is the biggest at LO and the smallest at NNLO. If we take two points, for example, 0.5 and 2, on the prediction line and take the difference between these points and default value, which is 1, we will have the biggest difference at LO due to having this biggest slope. Another remark should be underlined here that the predicted cross section value as a function of $\mu_R$ is not changing as much as it changes by $\mu_F$. Even LO QCD prediction does not show any reaction to the change of $\mu_R$ as discussed before.

The ratio of cross sections provided in Table 4 is highly dependent on factorization and renormalization scales. These numbers are extracted using HERAPDF2.0. Under normal conditions, an increase of cross section by increase of QCD orders is expected for most of the physics processes. That means the provided ratios in the table need to be above 1. However, $\sigma_{\text{NNLO}}/\sigma_{\text{NLO}}$ ratio drops below 1 in $\mu_R = \mu_F = XM_Z$ and $\mu_R = M_Z$, $\mu_F = XM_Z$ conditions. NLO prediction in the corresponding $X$ values will provide an overestimate of the NNLO cross section. It is also observed that LO predictions are getting greater than NNLO predictions at high $X$ points. There is a trend of decrease of $\sigma_{\text{NLO}}/\sigma_{\text{LO}}$ ratio in three conditions provided in Table 4. This is also true for $\sigma_{\text{NNLO}}/\sigma_{\text{NLO}}$ ratio in conditions of $\mu_R = \mu_F = XM_Z$ and $\mu_R = M_Z$, $\mu_F = XM_Z$. The same behavior is not observed on $\sigma_{\text{NNLO}}/\sigma_{\text{NLO}}$ while the condition is set to $\mu_F = M_Z$, $\mu_R = XM_Z$. All these outcomes draw a conclusion that wrong choices of renormalization and factorization scales may lead to bad perturbative behaviors.

Table 4: Predicted 14 TeV cross sections as function of $\mu_R$ and $\mu_F$. All numbers are in units of pb.

| X | $\sigma_{NNLO}$ | $\sigma_{NLO}$ | $\sigma_{LO}$ | $\sigma_{NNLO}/\sigma_{NLO}$ | $\sigma_{NLO}/\sigma_{LO}$ |
|---|---|---|---|---|---|
| | | $\mu_R = \mu_F = XM_Z$ | | | |
| 0.3 | 2137 | 2011.85 | 1372.98 | 1.062 | 1.465 |
| 0.6 | 2141 | 2073.73 | 1620.81 | 1.032 | 1.279 |
| 0.9 | 2143.4 | 2105.19 | 1757.64 | 1.018 | 1.198 |
| 1.3 | 2146.96 | 2132.4 | 1876.29 | 1.007 | 1.136 |
| 1.6 | 2150.03 | 2148.78 | 1940.9 | 1.001 | 1.107 |
| 1.9 | 2159.46 | 2163.87 | 1993.22 | 0.998 | 1.086 |
| 2.3 | 2165.9 | 2181.23 | 2050.03 | 0.993 | 1.064 |
| 2.6 | 2168.39 | 2188.52 | 2085.61 | 0.991 | 1.049 |
| 3.0 | 2172.35 | 2200.93 | 2126.15 | 0.987 | 1.035 |
| 3.3 | 2176.43 | 2210.13 | 2153.13 | 0.985 | 1.026 |
| 3.6 | 2185.69 | 2216.01 | 2176.86 | 0.986 | 1.017 |
| 4.0 | 2189.25 | 2228.18 | 2205.73 | 0.982 | 1.010 |
| | | $\mu_R = M_Z, \mu_F = XM_Z$ | | | |
| 0.3 | 2083.16 | 1906.25 | 1372.83 | 1.093 | 1.389 |
| 0.6 | 2131.04 | 2043.27 | 1620.75 | 1.043 | 1.261 |
| 0.9 | 2138.29 | 2100.86 | 1757.72 | 1.018 | 1.195 |
| 1.3 | 2142.32 | 2141.61 | 1876.37 | 1.000 | 1.141 |
| 1.6 | 2148.78 | 2160.52 | 1940.98 | 0.995 | 1.113 |
| 1.9 | 2149.76 | 2174.37 | 1993.3 | 0.989 | 1.091 |
| 2.3 | 2151.07 | 2187.3 | 2049.78 | 0.983 | 1.067 |
| 2.6 | 2150.51 | 2194.88 | 2085.36 | 0.980 | 1.053 |
| 3.0 | 2149.17 | 2202.7 | 2126.15 | 0.976 | 1.036 |
| 3.3 | 2147.89 | 2207.48 | 2152.87 | 0.973 | 1.025 |
| 3.6 | 2146.11 | 2211.36 | 2176.95 | 0.970 | 1.016 |
| 4.0 | 2145.11 | 2216.17 | 2205.73 | 0.968 | 1.005 |
| | | $\mu_F = M_Z, \mu_R = XM_Z$ | | | |
| 0.3 | 2195.04 | 2170.52 | 1792.26 | 1.011 | 1.211 |
| 0.6 | 2168.08 | 2134.94 | 1792.26 | 1.016 | 1.191 |
| 0.9 | 2148.45 | 2117.38 | 1792.26 | 1.015 | 1.181 |
| 1.3 | 2144.97 | 2103.91 | 1792.26 | 1.020 | 1.173 |
| 1.6 | 2144.97 | 2096.76 | 1792.26 | 1.023 | 1.170 |
| 1.9 | 2146.56 | 2090.69 | 1792.26 | 1.027 | 1.167 |
| 2.3 | 2143.74 | 2085.52 | 1792.26 | 1.028 | 1.164 |
| 2.6 | 2142.06 | 2081.57 | 1792.26 | 1.029 | 1.161 |
| 3.0 | 2139.47 | 2077.47 | 1792.26 | 1.030 | 1.159 |
| 3.3 | 2141.65 | 2074.82 | 1792.26 | 1.032 | 1.158 |
| 3.6 | 2139.49 | 2072.45 | 1792.26 | 1.032 | 1.156 |
| 4.0 | 2137.91 | 2069.66 | 1792.26 | 1.033 | 1.155 |

## 5. Conclusion

The main findings of these studies are reiterated here. First, total and fiducial cross section predictions at 13 TeV are compared with ATLAS study at 13 TeV to validate our code. It is observed that there is a good agreement between the predicted and observed results at 13 TeV and the comparisons showed that HERAPDF2.0 best describes the experimental results. The 14 TeV predictions are further computed by running the code used at 13 TeV comparison. 14 TeV fiducial cross section predictions are presented at NNLO QCD including NLO EW corrections for five different PDF models: HERAPDF2.0, MMHT2014, NNPDF3.0, CT14, and ABM12. All PDF predictions are in well agreement with each other. Total cross section predictions at 14 TeV are also reported using HERAPDF2.0 at LO, NLO, and NNLO QCD including

NLO EW corrections. In the calculations, PDF, $\alpha_s$, scale, model, and parameterization errors are considered. It is observed that the scale uncertainty is a dominant uncertainty at LO and NLO calculations and its contribution to the total uncertainty decreases from LO to NNLO.

Another topic studied here is the choice of $\mu_R$ and $\mu_F$ parameters. The obtained results show that the cross section increases by increase of $\mu_F$ and decreases by increase of $\mu_R$. Varying the two scales simultaneously leads to a compensation of the two different behaviors. As a result, the scale dependence of cross section is mostly driven by the factorization scale. These scale dependence characteristics are reduced when high-order corrections are included. Another important issue observed here is that NLO calculations at certain $X$ values provide an overestimate of the NNLO cross sections. The results also present a similar overestimation for LO predictions for high $X$ values. For certain renormalization and factorization values, LO predictions are higher than NNLO predictions although an increase of cross section from LO to NNLO is expected. $\mu_R$ and $\mu_F$ choices on the determination of scale uncertainty are further investigated. $\mu_R$ and $\mu_F$ scales are altered independently, where $\mu_R = XM_Z$ and $\mu_R = XM_Z$. Here $X$ could be 1, 1/2, and 2. We set 9 variations as listed $(\mu_R, \mu_F)$: (1, 1), (1, 2), (1, 1/2), (2, 1), (2, 2), (2, 1/2), (1/2, 1), (1/2, 2), and (1/2, 1/2). We found that there is very large uncertainty when $\mu_R/\mu_F = 4$. Based on the numbers presented in Table 1, it is found that the scale uncertainty is not altered significantly if the uncertainty at $\mu_R/\mu_F = 4$ is excluded. All these findings draw a conclusion that there is no right choice but smart choice; however, any wrong choice may lead to an unexpected perturbative behavior.

## Conflicts of Interest

The authors declare that they have no conflicts of interest.